\documentclass[twocolumn,pre]{revtex4}
\usepackage{epsfig}
\usepackage{amstext}
\begin{document}

\title{Stochastic Ratchet Mechanisms for Replacement of Proteins Bound to DNA}
\author{S. Cocco$^{1}$, J.F. Marko$^2$, R. Monasson$^{3}$}
\affiliation{
$^1$ Laboratoire de Physique Statistique de l'Ecole Normale Sup\'erieure, CNRS \& Universit\'e P.\&M. Curie, Paris, France\\
$^2$ Department of Molecular Biosciences and Department of
Physics and Astronomy,
Northwestern University, Evanston, Illinois 60208 \\
$^3$ Laboratoire de Physique Th\'eorique de l'Ecole Normale Sup\'erieure, CNRS \& Universit\'e P.\&M. Curie, Paris, France
}

\begin{abstract}
Experiments indicate that unbinding rates of proteins from DNA can depend on the concentration of proteins in nearby solution. 
Here we present a theory of multi-step replacement of DNA-bound proteins by solution-phase proteins.
For four different kinetic scenarios we calculate the dependence of protein unbinding and replacement rates on solution protein concentration.
We find  
(1) strong effects of progressive `rezipping' of the solution-phase protein onto DNA sites liberated by `unzipping' of the originally bound protein;
(2) that a model in which solution-phase proteins bind non-specifically to DNA can describe experiments on exchanges between the non specific DNA-binding proteins Fis-Fis and Fis-HU;  (3) that a binding specific model  describes experiments on the exchange of CueR proteins on specific binding sites.
\end{abstract}
\maketitle

Recent single-molecule experiments have revealed that exchange processes between proteins bound to DNA and proteins in solution can occur at rates in excess of spontaneous dissociation, increasing with protein solution concentration \cite{Gra11,Lop11,Jos12,Ha13}.  This effect is likely important to determining rates of turnover of proteins {\it in vivo}, where protein concentrations may be very high, and to understanding regulation of DNA transcription, replication, repair and packaging. Canonical models of protein-DNA interactions generally assume unbinding pathways dependent on a single rate-limiting step characteristic of the interaction of {\it one} protein molecule with its DNA substrate \cite{Dill}, and cannot explain this effect. In this paper, we introduce kinetic models aimed at describing the {\it multi-step} dynamics of bio-macromolecule interactions \cite{Kit69,noi}. 
This addition leads naturally to concentration-dependent {\it competition} between bound and `invading' molecules for substrate, and concentration-dependent replacement processes.
     
In \cite{Gra11} a single DNA was stretched out, and spontaneous dissociation of fluorescently-labeled   Fis (a bacterial chromosomal organization protein) was observed to be slow (about 90\% of initially bound protein remained bound for 30 minutes). However, when non-fluorescent protein solution  (either Fis or another DNA-binding protein, e.g., HU) was added, the fluorescent protein unbound rapidly (solution-phase Fis at 50 nM leads to 50\% dissociation after 3 minutes). The concentration dependence is described by a replacement (exchange) rate constant $R\simeq 6 \; 10^4\, {\rm M^{-1} s^{-1}}$ and $R\simeq 2.7 \; 10^3\, {\rm M^{-1} s^{-1}}$ for Fis-Fis and HU-Fis replacement. Ref. \cite{Jos12} studied a copper-ion (Cu$^+$) dependent transcription factor, CueR, which exists in a Cu$^+$-bound (CueR$^+$) and a Cu$^+$-free (CueR$^-$) conformations. Both conformations compete for a specific binding site to regulate genes protecting cells from copper-induced stress. Spontaneous dissociation of CueR$^+$ at a rate $k_{\rm off}=0.5\ {\rm s^{-1}}$  was sped up by CueR$^-$ in solution, with replacement rate $R\simeq 2.8\;10^7\ {\rm M^{-1} s^{-1}}$ (Fig.~4 of \cite{Jos12}).  In Ref. \cite{Lop11} it has been shown that fluorescently-labeled  polymerases in solution are recruited close to the replication fork, at a solution-concentration-dependent rate, and replace the DNA synthesizing polymerase at rate $k_{exc} \simeq 0.018\ {\rm s}^{-1}$.

These experiments show that off-rates of proteins from DNA can depend on solution-phase concentrations of proteins competing for the same DNA.
Here we describe a `stochastic ratchet' model of this competition. 
We propose that, due to thermal fluctuations, part of the bound protein releases from DNA, allowing a solution-phase protein to take its place. Rebinding of the released binding domain cannot occur because it requires thermal opening of the newly bound protein. Iterating this for a series of binding interactions allows gradual replacement. 
Through this process a solution-phase protein can replace a bound one far faster than if complete dissociation of the initially bound protein was required for its replacement.  Here, we introduce four distinct models of kinetic pathways for protein replacement.
We then compute concentration-dependent dissociation rates to determine which pathways best describe specific experiments.

\begin{figure}
\begin{center}
\epsfig{file=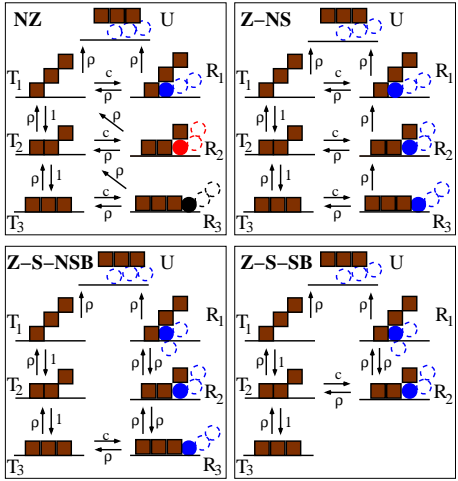,width=6cm}
\caption{ The four proposed unbinding pathways.
Brown squares show $N$ (=3 here) units of a protein bound to DNA (dark horizontal line). Circles show units of the invader proteins, with different colors corresponding to different proteins. Filled circles show units occupying the zipping site. The most likely replacement scenario at small concentration is shown with the blue invader protein. Parameters entering the rates are: 
(i) the mean number $c$ of solution-phase proteins per binding site, in units of the elementary concentration $c_o =1/ a^3$, of one particle per binding site, where $a$ is a length scale associated with the linear dimension of a binding site (for $a= 1$~nm, $c_o=1$~M);
(ii) the ratio of the unbinding and binding rates for one unit: $\rho=e^{-\epsilon}$, where  $\epsilon$ is the binding energy in $k_BT$ units. 
Time is expressed in terms of the time scale $t_o$, equal to the the self-diffusion time for one unit of the protein: $t_o =2 \pi \eta a^3/k_B T\approx 1.6\;10^{-9}$~s, for $a\approx1$~nm, $\eta=0.001$~Pa~s  and $k_B T= 4\; 10^{-21} $~J. In units of $1/t_0$ the zipping rate of a protein unit on a free binding site is equal to one.}
\label{models}
\end{center}
\end{figure}

Kinetic schemes for the four models are shown in Fig.~\ref{models}. We assume that binding of proteins to the DNA occurs by at most $N$ `units' ($N=3$ in Fig.~\ref{models}). Units bind to putative anchoring sites; values of $N$ and bond energy $\epsilon$ will be discussed below. Each model contains two unbinding pathways: a spontaneous unbinding pathway where the units of the bound protein dissociate one after the other with rate $\rho=e^{-\epsilon}$ (states T$_1$, ..., T$_N$ to the left of each box in Fig.~\ref{models}), and a replacement pathway (right) where the bound protein is replaced by invading proteins  (states R$_1$, ..., R$_N$). 
In the Non-Specific (NS) pathways, the invading protein may bind on each site left unoccupied by the bound protein, while in the Specific (S) pathways, binding occurs only from one extremity of the bound protein. In the Zipping models (Z) the invader, once attached to DNA, binds adjacent sites successively (zips) as they are released by the bound protein. 
In the Non-Zipping (NZ) model (top left panel, Fig.~\ref{models}), when one bound unit of the protein detaches, this site  is left unoccupied, corresponding to transitions R$_i \to$ T$_{i-1}$. By contrast, in the Z-models, as soon as the bound unit bound detaches, the invading protein occupies the vacant site, making transitions R$_i \to$ R$_{i-1}$ possible.

For the NZ and Zipping-Non-Specific (Z-NS) models (top panels of Fig.~\ref{models}), the invading proteins can bind with rate $c$ proportional to the concentration of solution proteins (transitions T$_i \to$ R$_i$), and unbind with rate $\rho$ (transitions R$_i \to$ T$_i$) on each possible anchoring site, {\em i.e.} whatever the value of $i$. 
We consider two Zipping Specific (Z-S) models (bottom panels of Fig.~\ref{models}).
In the Non-Specific Binding (Z-S-NSB) case, the invader may bind next to the fully DNA-bound protein (state R$_N$), which represents a non-specific nearby site.  
For the Specific Binding (Z-S-SB) case, the invader can only bind specifically to the first available site after one unit of the bound protein detaches (state R$_{N-1}$), after which zipping of the invader may then proceed.

The probability $P_S(t)$ that the system is in state $S$ at time $t$  is described by the master equation 
\begin{equation}
\frac{ d P_S(t)}{dt}=\sum_{S'}\; W(S \leftarrow S')\;P_{S'} (t)\ .
\end{equation}
The system is initially in state T$_N$, corresponding to a fully-bound protein, with no invader present, and eventually reaches the unbound state U (Fig.~\ref{models}). Off-diagonal, non-zero elements of the transition matrix $W$ are given for the four models in Fig.~\ref{models} and its caption (see also Supplementary Information). Diagonal elements conserve probability,
$W(S \leftarrow S)=-\sum_{ S'(\neq S)} W(S' \leftarrow S)$. 

\begin{figure}
\begin{center}
\epsfig{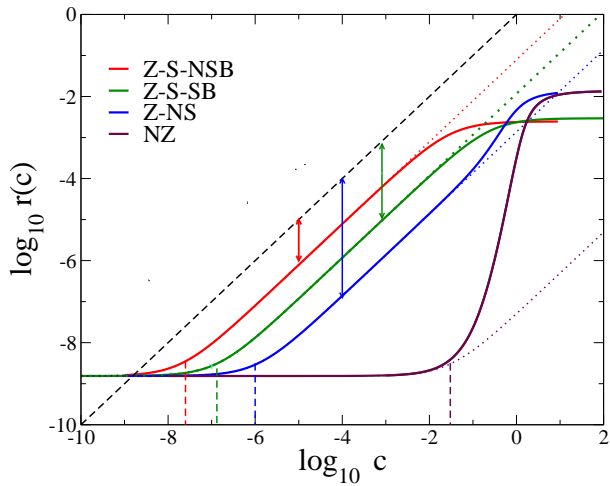}
\caption{Unbinding rates $r(c)$ (full lines) from Eq.~(\ref{avtte}) versus dimensionless concentration $c$ and for parameters $N=10$, $\epsilon=2$. The zero-concentration rate, $r(0)=1.8\; 10^{-9}$ (units of $1/t_0$), is the same for all four models. The concentrations $c_R$ at which replacement starts to dominate over pure thermal unbinding, {\em i.e.} $r(c)$ starts to vary linearly with $c$, are indicated  by the vertical dashed colored lines for the four models; $c_R= 3\;10^{-2}$ for NZ, $2\;10^{-6}$ for Z-NS, $2\,10^{-8}$ for Z-S-NSB, and  $1.3\;10^{-7}$ for Z-S-SB (units of $c_0$). The offsets between the linear regimes of the rate curves (dotted lines, from Eq.~(\ref{pt0})) and the $r=c$ (dashed black) line are $\log R$ (log of replacement rate, double arrow vertical lines); $R$ is approximatively equal to $10^{-1}$  for Z-S-NSB,  $10^{-2}$  for Z-S-SB and $10^{-3}$ for Z-NS.
 The dotted lines are only visible when the linear approximation breaks down. }
\label{logratevslogcompmodels}
\end{center}
\end{figure}

Given the $W$ matrix, the average occupancy time, or equivalently the inverse of the binding rate $r(c)$ of the protein, is
\begin{eqnarray}
\frac 1 {r(c)}=\int_{0}^{\infty}dt\sum _{S\ne U}P_S(t)  = - \langle B| W^{-1}|T_N\rangle \ ,
\label{avtte}
\end{eqnarray}
where $W^{-1}$ is the inverse matrix of $W$, and $|B\rangle$ denotes the sum of all $2N$ bound states $|$T$_i\rangle$ and $|$R$_i\rangle$, with $1\le i\le N$. The unbinding rate $r(c)$ is plotted as a function of $c$ in Fig.~\ref{logratevslogcompmodels} for the four models of Fig.~\ref{models}. 

Without solution-phase protein ($c=0$) the unbinding rate is $r(0)=(1-\rho)^2 \rho^N$ (up to $O(\rho^{2N})$ corrections) for all four models: pure thermal unbinding is exponentially slow in $N$. The unbinding rate at small concentration $c>0$ can be studied perturbatively. Using the linear dependence of $W$ on $c$ we write $W=W_o + c\;W_1$, where spontaneous dissociation is described by $W_o$ and invasion-zipping is described by $W_1$. We have $W^{-1}(c)=W_o^{-1} - c\;W_o^{-1}\; W_1\; W_o^{-1}+ 0(c^2)$. The mean unbinding rate is therefore approximately 
\begin{equation}
r(c) \simeq r(0) + R \, c +O(c^2)\ ,
\label{pt0}
\end{equation}
where the replacement rate $R$ is
\begin{equation}
R\equiv\left.\frac{dr}{dc}\right|_{c=0} =-r(0)^2 \;\langle B|W_o^{-1}\; W_1\; W_o^{-1}|T_N\rangle \ .
 \label{pt}
\end{equation}
We define the replacement concentration $c_R$ as the concentration 
at which the unbinding rate is twice its zero-concentration value,
\begin{equation}
c_R= \frac{r(0)}{R}\ .
\end{equation}
The unbinding rate increase at small concentration, $R\,c$, is the rate at which unbinding-replacement involves essentially one invading protein. The most likely unbinding scenario is indicated by the sequence of blue invader configurations in Fig.~\ref{models}, providing an approximation sufficient to understand the scaling of the replacement rate $R$ with the number of binding units, $N$ (see Supplementary Information for details).

For the NZ model (brown curve in Fig.~\ref{logratevslogcompmodels}) we find
a replacement rate, Eq.~(\ref{pt}), of
\begin{equation}
R^{NZ}=\rho^{N-1} \frac{(N-1-N \,\rho-\rho) \;(1-\rho)}{2}+ O(\rho^{2N-1})\ .
\label{replmod1}
\end{equation}
The most probable unbinding scenario with replacement at small $c$ is that, from state T$_1$, occupied with probability $\rho^{N-1}$, the invader binds with rate $c$  (transition T$_1\to$ R$_1$ in Fig.~\ref{models}, top \& left). The protein is then equally likely to dissociate (R$_1 \to$ U) or to come back to the thermal pathway (R$_1\to$ T$_1$). We therefore obtain $R^{NZ} \propto \rho^{N-1}$, in agreement with (\ref{replmod1}). As $R^{NZ}$ is exponentially small in $N$, the replacement concentration $c^{NZ}_R\sim \frac{2 \rho}{N-1}$ can become large; for a binding energy $\epsilon=2~k_BT$, $N=10$ protein units, and $c_0=1$~M,  we find $c^{NZ}_R= 3\;10 ^{-2}$~M (Fig. \ref{logratevslogcompmodels}), well above that experimentally observed (tens of  nM \cite{Gra11, Jos12}). 
The concentration range where the linear approximation, Eq.~(\ref{pt0}), holds is very narrow. Contrary to experiments, the unbinding rate of the NZ model shows a highly nonlinear concentration-dependence, $r(c)\sim c^N$ for $c\gtrsim c_R$ (Fig.~\ref{logratevslogcompmodels}).

For the Z-NS model (blue curve in Fig.~\ref{logratevslogcompmodels}) the replacement rate,
\begin{equation}\label{replmod2}
R^{Z-NS}=\frac{(1-\rho)^2}{(1-2\,\rho)^2 2^N} +O(\rho^{N-2}) \ ,
\end{equation}
is also exponentially small in $N$, but decays less quickly with $N$ than in the NZ model, as $\rho$ is generally smaller than $\frac 12$. In the most likely unbinding scenario (blue configurations, Fig.~\ref{models}, top \& right), the invader attaches through the transition T$_N\to$ R$_N$ from the frequently occupied, fully-bound state T$_N$. The scenario follows the replacement pathway all the way up with probability $2^{-N}$, until the protein is released and gives 
the scaling $R^{NZ} \propto 2^{-N}$ (\ref{replmod2}).
Figure~\ref{logratevslogcompmodels} shows that the linear approximation (\ref{pt0},\ref{replmod2}) describes the exact unbinding rate $r(c)$  (blue curve; see Supplementary Information for results obtained for varied $N$ and $\epsilon$). 
At large concentrations the unbinding rate grows $\approx c^N$ as in the NZ model, since invading proteins can attach and attempt replacement at every site. 

\begin{figure}
\begin{center}
\epsfig{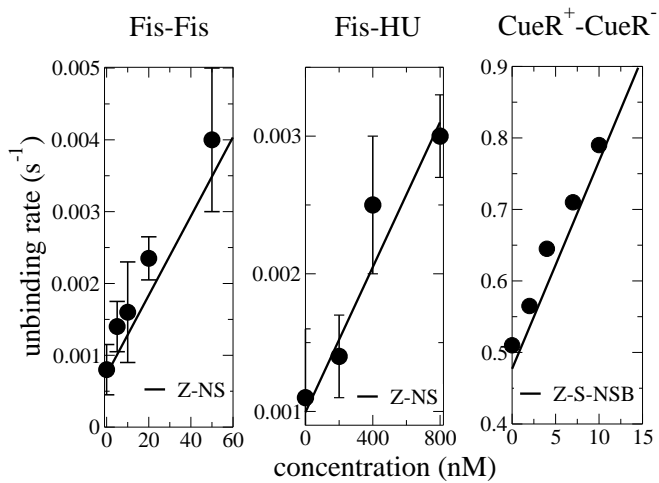}
\caption{Fit of concentration-dependent unbinding rates of Fis bound to DNA in the presence of Fis (left) and HU (middle) proteins in solution \cite{Gra11}, and for CueR dissociation rates as a function of CueR concentration in solution (right) \cite{Jos12}, using $a=1$~nm \cite{note}, and N=14, $\epsilon=1.95$ for Fis-Fis,
N=19, $\epsilon=1.4$ for Fis-HU, and N=15 $\epsilon=1.36$ for CueR-CueR.
For replacement rates (slopes of unbinding rates versus concentration) we find
  $R^{Fis-Fis}= 5\;10^4$~M$^{-1}\,$s$^{-1}$, $R^{Fis-HU}=2.6\;10^3$~M$^{-1}\,$s$^{-1}$, and $R^{CueR-CueR}= 2.9\; 10^7$~M$^{-1}\,$s$^{-1}$, in agreement, considering the error bars, with experimental fits. Replacement concentrations found with the Z-NS model are $c_R= 2$~nM for Fis-Fis and $c_R= 370$~nM for Fis-HU, while we find $c_R=16$~nM for CueR-CueR with the Z-S-NSB model. }
\label{ratevsc}
\end{center}
\end{figure}

Figure~\ref{ratevsc} compares the Z-NS model to experiment for Fis-Fis and Fis-HU replacement. 
We fit $N$ and $\epsilon$ (see Fig.~\ref{ratevsc} caption)
by matching experimental exchange and spontaneous unbinding rates 
to  $R^{Z-NS}$ in Eq.~(\ref{replmod2}) and $r(0)$. Fis-Fis replacement dominates thermal unbinding as soon as $c\gtrsim c_R\simeq (2\rho)^N\simeq 2$~nM, a value of concentration compatible with experimental observation~\cite{note}. 
Heterotypic Fis-HU replacement dominates at concentrations of hundreds of nM \cite{Gra11}, in agreement with experiment. 
In both cases the binding energy $\epsilon$ is of the order of $\simeq 2~k_BT$. However, for CueR$^+$-CueR$^-$ data \cite{Jos12} we obtain a binding energy per site of $\epsilon \simeq 5~k_BT$, 
which is too large even for specific binding. 
The Z-NS model can reasonably decribe Fis-Fis and Fis-HU exchange, 
but not CueR$^+$-CueR$^-$ exchange.

For DNA polymerase \cite{Lop11} the observed exchange rate following recruitment is very small, $\approx 2\times 10^{-2}$~s$^{-1}$, with no concentration dependence since the polymerase is recruited and exchanged in separate kinetic steps. 
However, we can still proceed by setting the dimensionless concentration to $c=1$. 
Using the Z-NS model with $a=1$~nm we have $t_o=1.6\,10^{-9} s$  which  gives $N=35$. 
This large $N$ value is consistent with the large DNA-binding surface of DNApol. 

The replacement rate for the Z-S-NSB model is
\begin{equation}\label{replmod3}
R^{Z-S-NSB} =  \frac {1-\rho} {N+1} +O(\rho^{N-1}) \ .
\end{equation}
It decreases only algebraically with $N$, 
and is much larger than its Z-NS counterpart which scales as $2^{-N}$. 
In the most probable replacement pathway the invader attaches at the last zipping site 
(T$_N$ in Fig.~\ref{models}, bottom \& left) with rate $c$.  
The probability that the system continues along the replacement pathway until the bound protein is released, and never reaches T$_N$ again, scales as  $1/(N+1)$, 
giving the $R^{Z-S-NSB}$ scaling in (\ref{replmod3}). 
The linear approximation for $r(c)$ (\ref{pt0}, \ref{replmod3}) is valid 
over a large range of concentration (red curve in Fig.~\ref {logratevslogcompmodels}). 
The Z-S-NSB model allows us to fit the replacement rate of the CueR$^+$-CueR$^-$ experiment, which is about $1000$ times larger than the one observed for Fis-Fis replacement, and the corresponding replacement concentration $c_R\simeq (N+1)\rho^N$, with reasonable parameter values $N=13-16$ and $\epsilon=1.2-1.4$, giving a total binding energy of the order of tenth of $k_B T$ (Fig.~\ref{ratevsc}). 
The Z-NS model is inappropriate to describe CueR$^+$-CueR$^-$  replacement, 
as it requires $N\sim 10^4$ to generate the observed exchange rate.
For the Z-S-SB model the replacement rate reads
\begin{equation}
\label{replmod6}
R^{Z-S-SB} =  \frac {\rho(1-\rho)} {N}  +O(\rho^N) \ ,
\end{equation}
with replacement concentration $c_R\simeq N\rho^{N-1}$. 
The scaling with $N$  is similar to that of Z-S-NSB, 
with $1/N$ instead of $1/(N+1)$ due to the shorter replacement pathway (Fig.~\ref{models}). 
The Z-S-SB model does not reproduce the CueR$^+$-CueR$^-$ replacement rate. 
Indeed the replacement rate is smaller by a factor $e^{-\epsilon}\simeq 3-4$ with respect to the rate obtained with the Z-S-NSB model for $\epsilon=1-1.4$~k$_B T$  and $N$ ranging from 13 to 16 as in Fig.~\ref{logratevslogcompmodels} (green curve). 
 
In conclusion we have introduced four kinetic models to describe replacement processes between DNA-bound proteins. We have solved the models 
at three  levels: numerically, using a small--concentration expansion providing analytical formulae for the replacement rate  $(R_c),$  and with a 'dominant pathway' approximation which gives the correct scaling of $R_c$ with $N$.  
The importance of zipping for efficient bound-invader exchange is illustrated by the inability of the NZ model to replace bound proteins at low concentrations. The Z-NS model is appropriate to describe Fis-Fis  and Fis-HU exchanges, presumably due to the non-specific nature of Fis- and HU-DNA interactions \cite{Gra11}. Moreover the Z-NS model is appropriate to describe polymerase replacements, since DNA-pol is able to interact with any DNA sequence. 
Z-S-NSB better describes CueR-CueR exchange, which is sensible since CueR interacts with a specific binding site \cite{Jos12}.  
Fits for $N$ range from 14 to 35 depending on the protein, and the fit values of binding energy per unit are 
$\approx\epsilon=1-2 k_BT$, consistent with individual non-covalent interactions. 
DNA-binding proteins interact in complex ways with their substrate \cite{rhos2010}, and the number $N$ of `units' used here can be thought of as an effective number of separate bonds.  For the heterogeneous Fis-HU replacement  the fit value of $\epsilon$ is smaller than for homogeneous Fis-Fis replacement; this is consistent with the larger DNA binding affinity of Fis compared to HU \cite{Gra11}.  A more general (and precise) model would describe the invading protein through a binding energy $\epsilon'$ different from the binding energy $\epsilon$ for the bound protein in case of heterogeneous replacement, as well through its number of bound units, rather than the on/off description used here. 
Binding/unbinding of small DNA fragments (oligonucleotide) on a DNA under force \cite{Din12} and exchange of DNA-binding oligonucleotides in DNA hybridization assays \cite{Ott14,carlon,Dup13} are likely described by the Z-S-SB model.  Including sequence specificity (dependence of $\epsilon$ on the sites) could help in modeling such experiments \cite{Din13}.


While revising this article two papers have appeared providing further evidence for protein exchange \cite{Gib13,Luo13}, suggesting the generality of the replacement process. 
Luo and collaborators \cite{Luo13} have seen displacement of transcription factors 
by nucleosomes on DNA; see  \cite{Che14} for a related theory.  
Gibb and collaborators \cite{Gib13} have demonstrated exchange of replication protein A (RPA) and the recombinase Rad51 and exchange of RPA with single-stranded-binding (SSB) protein on single-stranded DNA; the kinetic scheme in Fig.~6 of \cite{Gib13} corresponds to our Z-NS model.

{\bf Acknowledgement:} We thanks V. Croquette for useful discussions. J.F. Marko  acknowledges NSF Grants MCB-1022117 and DMR-1206868, and  NIH Grant 1U54CA143869-01 (NU-PS-OC).

\end{document}